\documentclass[journal,10pt]{IEEEtran}
\usepackage{xcolor}
\usepackage{relsize}

\usepackage{ifpdf}
\usepackage{graphicx}
\usepackage{cite}
\usepackage{nomencl}
\usepackage{enumitem}
\usepackage{slashbox}
\usepackage{bm}
\usepackage[normalem]{ulem}

\ifCLASSINFOpdf
\else
\fi
\usepackage{amsmath,amssymb,bm}
\usepackage{algorithmic}
\usepackage{adjustbox}
\usepackage{array}
\usepackage[tablename=Table]{caption}
\usepackage{stfloats}
\usepackage{url}
\usepackage{amsthm}
\usepackage{multirow}
\usepackage{algorithm}

\newcommand{\RomanNumeralCaps}[1]{\MakeUppercase{\romannumeral #1}}
\usepackage{stackengine,scalerel}

\newcommand\overstar[1]{\ThisStyle{\ensurestackMath{%
			\setbox0=\hbox{$\SavedStyle#1$}%
			\stackengine{0pt}{\copy0}{\kern.2\ht0\smash{\SavedStyle*}}{O}{c}{F}{T}{S}}}}

\newcommand{\tr}[0]{ \text T}
\newcolumntype{P}[1]{>{\centering\arraybackslash}p{#1}}

\DeclareMathOperator*{\argminA}{arg\,min}
\DeclareMathOperator*{\argmaxA}{arg\,max}
\begin{document}
 
\title{An Information Theoretic approach to identify Dominant Voltage Influencers for Unbalanced Distribution Systems}


\author{\IEEEauthorblockN{Sai Munikoti,~\textit{Student Member, IEEE}, Mohammad Abujubbeh,~\textit{Student Member, IEEE}, \\ Kumarsinh Jhala,~\textit{Member, IEEE},  Balasubramaniam Natarajan,~\textit{Senior Member, IEEE} 
		\thanks{K. Jhala is with the Center for Energy, Environmental, and Economic Systems Analysis in the Energy Systems Division at Argonne National Laboratory (e-mail: kjhala@anl.gov).}
		\thanks{S. Munikoti, M. Abujubbeh and B. Natarajan are with Electrical and Computer Engineering, Kansas State University, Manhattan, KS-66506, USA, (e-mail: saimunikoti@ksu.edu, abujubbeh@ksu.edu, bala@ksu.edu)}
		\thanks{This work has been submitted to the IEEE for possible publication. Copyright may be transferred without notice, after which this version may no longer be accessible.}}}


\maketitle

\begin{abstract}
Smart distribution grid with multiple renewable energy sources can experience random voltage fluctuations due to variable generation, which may result in voltage violations. Traditional voltage control algorithms are inadequate to handle fast voltage variations. Therefore, new dynamic control methods are being developed that can significantly benefit from the knowledge of dominant voltage influencer (DVI) nodes. DVI nodes for a particular node of interest refer to nodes that have a relatively high impact on the voltage fluctuations at that node.
Conventional power flow-based algorithms to identify DVI nodes are computationally complex, which limits their use in real-time applications. This paper proposes a novel information theoretic voltage influencing score (VIS) that quantifies the voltage influencing capacity of nodes with DERs/active loads in a three phase unbalanced distribution system. VIS is then employed to rank the nodes and identify the DVI set. VIS is derived analytically in a computationally efficient manner and its efficacy to identify DVI nodes is validated using the IEEE 37-node test system. It is shown through experiments that KL divergence and Bhattacharyya distance are effective indicators of DVI nodes with an identifying accuracy of more than $90$\%. The computation burden is also reduced by an order of $5$, thus providing the foundation for efficient voltage control.

\end{abstract}


\IEEEpeerreviewmaketitle
\section{Introduction}
\IEEEPARstart{T}{he} power grid is evolving with the increasing integration of renewable distributed energy resources (DERs). While offering the benefits of reduced carbon footprint, DERs also impose new technical challenges. Specifically, active consumers with rooftop photovoltaics and distributed generation are expected to alter their generation and usage patterns to follow the trends of time-varying electricity prices \cite{jhala2017real}. This in turn induces frequent power variations. The uncontrolled operations of DERs under this condition leads to voltage fluctuations in the distribution system. Grid operators have to manage DERs' operations to assure the reliability of the grid. Therefore, various control algorithms have been considered for regulating power injection/consumption across various buses of the network to mitigate voltage fluctuations \cite{kulmala2014coordinated, barr2014integration,tao2018voltage}. However, traditional methods of voltage control involving voltage regulators, on-load tap changing transformers are slow and inadequate to deal with bi-directional power flows and fast dynamics. This necessitates the development of computationally efficient and fast dynamic voltage control algorithms that can handle the dynamics of power/voltage variation \cite{ccimen2018mitigation, acharya2018control, jhala2019data}. The speed of response to impending voltage issues and the computational efficiency of these dynamic algorithms rely on our ability to select the optimal set of nodes for control that have the highest influence on the system voltage profile.

To this end, we introduce the notion of dominant voltage influencer (DVI) nodes. Throughout this paper, we refer to \textit{actor nodes} as the nodes where power changes (due to DER unit and/or load dynamics) and \textit{observation nodes} represent the set of nodes where voltage state is monitored. Thus, DVI nodes for a particular observation node denote all those actor nodes that have a relatively high impact on the voltage state of that observation node compared to the rest of the actor nodes. Hence, altering the actions of DVI nodes result in the highest reduction in voltage issues at the observation node. The nodes could be DVI because of their size (generation/load capacity) and/or locations in the distribution network. 
Conventional methods of identifying such DVI nodes involve Monte-Carlo simulations using load flow algorithms. These classical methods possess various drawbacks including (1) high computational complexity, (2) numerical results with no analytical insights, and (3) scenario dependent results with no generalization. These factors limit the applicability of conventional approaches in modern distribution systems. Thus, there is a need to develop an analytical and computationally efficient algorithm for identifying DVI nodes in three phase unbalanced distribution systems, which in turn can be used to develop effective voltage control algorithms.

For identifying DVI nodes, we need to quantify the impact of actor nodes on other nodes of the network in terms of voltage. Therefore, first, we develop a novel metric namely voltage influence score (VIS) for quantifying the voltage influencing capacity of an actor node on any arbitrary observation node. The VIS is based on an analytical method of voltage sensitivity. Thereafter, the VIS is employed to rank the actor nodes for any arbitrary observation node and identify the most influential ones, i.e., DVI nodes. The proposed approach allows us to identify DVI nodes without relying on computationally expensive Monte-Carlo simulations thereby, significantly reducing the computation time.

\textbf{Related work: }
Various control algorithms are proposed in the literature for regulating voltage fluctuations in the distribution system. Primarily, the control algorithms can be broadly categorized into two main categories: (1) centralized \cite{barr2014integration,brenna2012automatic,kulmala2014coordinated,elkhatib2011novel}, and (2) decentralized \cite{tao2018voltage,demirok2011local,tonkoski2010coordinated} control approaches. Centralized voltage control algorithms are generally based on classical power flow solution, which can be computationally expensive for real time applications. On the other hand, decentralized voltage control methods trade-off performance for reduced computational complexity. The speed, complexity and performance of both centralized and decentralized algorithms rely on the size and space of control actions. In modern distribution systems, the change in power consumption or injections due to multiple active consumers necessitates the need for analyzing a large number of scenarios, expanding the size and space for determining optimal control actions. This may further increase the computational complexity of traditional voltage control algorithms. 
Other approaches to bring voltage to safe operational limits in distribution systems involve the installation of additional control devices such as, static-var compensators \cite{parniani1995voltage, wan2018voltage}, dynamic voltage restorers \cite{haque2001compensation}, and grid-edge volt-var controllers \cite{moghe2016grid}. Despite their effectiveness in voltage control, the installation of additional devices in the system can be expensive. Therefore, a practical and economical solution to regulate voltage fluctuations is to utilize smart grid edge devices like PV inverters.
To this end, multiple dynamic algorithms have been proposed to control inverters and loads to regulate voltages \cite{ccimen2018mitigation, acharya2018control, jhala2019data}. In \cite{ccimen2018mitigation}, authors randomly select and shut down the thermostatically controlled loads during the time of high voltage unbalance. Similarly, \cite{jhala2019data} proposes a data-driven voltage control algorithm to regulate voltage violations by minimizing power injections at nodes with over voltage violations. In a nutshell, all these prior works use heuristic based methods to identify nodes whose generations need to be curtailed. However, such irrational selection is not desirable in terms of yielding optimal solution and providing fairness to all generators participating in a transactive energy market. 

Furthermore, there are some efforts where system is partitioned into smaller clusters and control is implemented for each cluster in a distributed manner \cite{bahramipanah2015network,bahramipanah2016decentralized,mahmoud2020three,zhao2017network}. In \cite{bahramipanah2015network,bahramipanah2016decentralized}, for instance, authors propose a new clustering method based on voltage sensitivity with respect to power fluctuation due to battery storage systems (BSS). However, this method is scenario specific since the location of BSSs in the system is fixed. Another sensitivity based zonal voltage control method is proposed in \cite{mahmoud2020three} where all nodes in a particular cluster are assumed to participate in power curtailment for voltage control. This is difficult to implement as it is not economical or even feasible for utilities to have access to control all nodes. 
There are some approaches which also include topological data for system clustering\cite{cotilla2013multi,zhao2017network,baranwal2019clustering}. For the approach in \cite{zhao2017network} to work, the nodal Q-V sensitivities must be computed to iteratively evaluate the modality index for each cluster \cite{girvan2002community}. Nevertheless, the computational complexity of this method increases with the size of the system as computing the Q-V based modality index of each cluster is based on forming the adjacency matrix of the system. Additionally, the entire body of prior work on identifying dominant nodes of voltage fluctuations does not consider stochasticity associated with DER power injections or user load variations. 

To address these demerits of existing approaches, analytical and computationally efficient approaches to identify DVI nodes that have relatively high influence on the voltage state of critical nodes are needed. The work in this paper addresses this essential research gap using a new information theoretic voltage influencing index that is applicable to three-phase unbalanced systems. \\
\textbf{Contributions:}
This paper proposes a novel approach to identify DVI nodes that can be used for network clustering and distributed control mechanisms in unbalanced distribution systems with DERs. The key contributions of this paper include: 
\begin{itemize}
    \item An analytical and computationally efficient method of identifying dominant voltage influencer nodes is proposed for three phase unbalanced distribution system.
    \item This work introduces a VIS metric that quantifies the voltage influencing capacity of nodes with DERs/active loads, and a computationally efficient method to compute it.
    \item Compared to \cite{jhala2019dominant}, this work has three major advancements (1) the probabilistic model of voltage fluctuations is derived for a three phase unbalanced distribution systems; (2) The voltage influencing capacity is quantified for each pair of nodes; (3) In addition to identifying the most dominant node, this work provides ranks of all dominant influencer nodes based on their VIS;    
    \item The effectiveness of proposed method is evaluated on the standard unbalanced IEEE 37-node test system yielding an accuracy of more than $90$\% with the computation complexity reduced by an order of $5$ compared to a classic load flow based approach.
\end{itemize}

The paper is organized as follows. Section \RomanNumeralCaps{2} presents a conventional method to identify DVI nodes followed by an analytical method of voltage sensitivity analysis. Section \RomanNumeralCaps{3} computes probability distributions of voltage change. Various information theoretic
indicators of the DVI nodes are proposed in Section \RomanNumeralCaps{3}, which are used to compute VIS. The performance of the proposed approach is tested using the IEEE 37-node test system in Section \RomanNumeralCaps{4}. Conclusions and future-work are presented in Section \RomanNumeralCaps{5}.
\vspace{-0.2cm}
\section{Background}
This section introduces the conventional approach to identify Dominant voltage influencer (DVI) nodes in a power distribution network. Then, the fundamental analytical expressions of voltage sensitivity are discussed which are later used to model voltage fluctuations for identifying DVI nodes. 
\vspace{-0.3cm}
\subsection{Conventional approach to identify DVI nodes}
The DVI nodes for an observation node are the nodes that have high impact on the voltage fluctuations at the observation node. An actor node can be a DVI due to association of two factors: (1) location of the actor node, i.e. phase and bus of the distribution network, and (2) generation/load capacity of DER/loads connected at the actor node which influences the variance of power change at that node. Generally, simulation-based scenario analysis is used as a major planning tool to identify DVI nodes. A typical approach involves following steps \cite{jhala2019dominant}: 
\subsubsection{Step 1- Compute variance of voltage change at each phase of observation node due to all actor nodes} The variance of voltage change at each phase of the observation node is computed by running multiple power flow based Monte-Carlo simulations with varying power, which captures temporal variation of generations.  
\subsubsection{Step 2- Calculate reduction in variance of voltage change at the observation node due to each actor node by setting power drawn/injected by the actor node as zero} This step requires repetition of Step 1 for each actor node after setting the variance of actor node as zero. 
\subsubsection{Step 3- Rank actor nodes based on the reduction in variance caused by the removal of the corresponding actor node:} Actor nodes are ranked in an ascending order with topmost and bottommost rank assigned to those actor node that causes maximum and minimum reduction in variance of voltage change at the observation node, respectively.

The scenario-based simulations incur high computational complexity, which grows with the size of the network. Therefore, various information theoretic metrics are explored in the next section that can identify DVI nodes in a computationally efficient manner, and can be used as a tool in various power system operations.
\begin{algorithm}[h!]
 \caption{Proposed method to identify DVI nodes}
 \begin{algorithmic}[1]
 \renewcommand{\algorithmicrequire}{\textbf{Input:}}
 \renewcommand{\algorithmicensure}{\textbf{Output:}}
 \REQUIRE Distribution Network with the target observation node. 
 \ENSURE  Ranked actor nodes for the input observation node. 
  \STATE  Compute distribution of voltage change at the input observation node due to each actor node at a time.
  \STATE  Compute distribution of voltage change at the input observation node due to aggregate effect of all actor nodes.
 \STATE Find statistical distance between the distributions computed in steps 1 \& 2.
 \STATE Compute VIS for each pair of observation and actor nodes.
 \STATE Rank actor nodes in the ascending order based on their respective VIS.
 \RETURN ranks and VIS of actor nodes for an input observation node.
 \end{algorithmic} 
 \end{algorithm}
 \vspace{-0.9cm}
\subsection{Analytical VSA}
In a three phase distribution system, change in power at any one phase of the node causes change in voltage at all phases of all the other nodes. Traditional methods of voltage sensitivity analysis (VSA) are computationally complex and less generic. Therefore, we developed a computationally efficient method of VSA in \cite{munikoti2020analytical,munikoti2020probabilistic}. For an unbalanced power distribution system, change in complex voltage $\Delta V_{O}$ at an observation node ($O$) due to change in complex power at multiple actor nodes can be approximated as:
	\begin{equation}	
	\begin{bmatrix}
	\Delta V_{O}^{a} \\[15pt]
	\Delta V_{O}^{b} \\[15pt]
	\Delta V_{O}^{c} 
	\end{bmatrix} \approx- \sum_{A \epsilon \tilde{A}}  \left(	
	\begin{bmatrix}
\frac{\Delta S_{A}^{a\star}Z_{OA}^{aa}}{ V_{A}^{a\star}} + \frac{\Delta S_{A}^{ b\star}Z_{OA}^{ab}}{ V_{A}^{b\star}}+ \frac{\Delta S_{A}^{ c \star}Z_{OA}^{ac}}{ V_{A}^{c\star}} \\[8pt]
\frac{\Delta S_{A}^{ a\star}Z_{OA}^{ba}}{ V_{A}^{a\star}} + \frac{\Delta S_{A}^{ b\star}Z_{OA}^{bb}}{ V_{A}^{b\star}} + \frac{\Delta S_{A }^{c\star}Z_{OA}^{bc}}{ V_{A}^{c\star}}  \\[8pt]
\frac{\Delta S_{A}^{ a\star}Z_{OA}^{ca}}{ V_{A}^{a\star}} + \frac{\Delta S_{A}^{ b\star}Z_{OA}^{cb}}{ V_{A}^{b\star}}+ \frac{\Delta S_{A}^{ c\star}Z_{OA}^{cc}}{ V_{A}^{c\star}}
\end{bmatrix}\right),
\label{eq:1} 
\end{equation}
where $a,b$ and $c$ represent the three phases. $V_{A}^{a}$ and $\Delta S_{A}^{a}$ represent complex voltage and power changes at the phase $a$ of the actor node $A$, respectively; $Z$ denotes the impedance matrix including self and mutual line impedance of the shared path between observation node and actor node from the source node. $\tilde{A}$ is the set of all actor nodes. Eqn. (\ref{eq:1}) follows superposition law and voltage change is aggregated due to the power change of each actor nodes. For a single actor node, the voltage change at any phase (say phase $a$) of an observation node $O$ can be decomposed into real ($\Delta V_{OA}^{a,r}$) and imaginary components ($\Delta V_{OA}^{a,i}$) as:

\begin{equation}
\begin{split}
\pmb{\Delta V_{OA}^{a,r}} \approx  \Bigg[ -
\frac{(\Delta P_{A}^{a}R_{OA}^{aa}+\Delta Q_{A}^{a}X_{OA}^{aa})(V_{A }^{a,r})}{(V_{A}^{a,r})^{2}+(V_{A}^{a,i})^{2}} + \\ 
\frac{(\Delta P_{A}^{a}X_{OA}^{aa}-\Delta Q_{A}^{a}R_{OA}^{aa})(V_{A}^{a,i} )}{(V_{A}^{a,r})^{2}+(V_{A}^{a,i})^{2}}- \dots  \Bigg] ,\\[6pt]
\end{split}
\label{eq:10b}
\end{equation}

\begin{equation}
\begin{split}
\pmb{\Delta V_{OA}^{a,i}} \approx \Bigg[-
\frac{(\Delta P_{A}^{a}X_{OA}^{aa}-\Delta Q_{A}^{a}R_{OA}^{aa})(V_{A }^{a,r})}{(V_{A}^{a,r})^{2}+(V_{A}^{a,i})^{2}} - \\ 
\frac{(\Delta P_{A}^{a}R_{OA}^{aa}+\Delta Q_{A}^{a}X_{OA}^{aa})(V_{A}^{a,i} )}{(V_{A}^{a,r})^{2}+(V_{A}^{a,i})^{2}}- \dots \Bigg] ,\\[6pt]
\end{split}
\label{eq:10b}
\end{equation}
where $\Delta P_{A}^{a}$ and $\Delta Q_{A}^{a}$ are the real and reactive power change at phase-$a$ of the actor node $A$, $R_{OA}^{aa}$ and $X_{OA}^{aa}$ are resistance and reactance of shared line between observation node $O$ and actor node $A$.
This work leverages this analytical formulation of VSA to derive voltage change distributions and consequently identify the DVI nodes in a three-phase unbalanced distribution system.

\section{Proposed framework }
As can be seen from the previous section, the conventional method of identifying DVI nodes is computationally complex, which limits its use in real-time operation of power system. Therefore, we propose the use of information-theoretic distance metrics as potential indicators of DVI nodes. Fundamentally, the proposed approach consists of four steps. In the first step, we obtain  probability distributions of voltage change at an observation node due to each actor node as well as due to the aggregate presence of all actor nodes. These distributions are derived in a computationally efficient way by employing the analytical expressions discussed in the background section. In the second step, we compute distances between the probability distribution due to each actor node and the distribution due to the aggregate effect of all actor nodes. The Third step utilizes the distances to compute voltage influencing score (VIS) for each pair of observation and actor nodes. Finally, for each observation node, all the actor nodes are ranked based on the computed VIS. The actor node whose voltage change distribution is nearest to the aggregate voltage change distribution is deduced as the major influencer of voltage change for that particular observation node. Likewise, all actor nodes are ranked based on the ascending order of their distances. The complete procedure is summarized in the Algorithm $1$. 
\vspace{-0.2cm}
\subsection{Probabilistic model of voltage fluctuations}
This section provides a probabilistic model of voltage change in a three-phase unbalanced distribution system, which consist of multiple spatially distributed actor nodes with PVs and active consumers. Random change in power at actor nodes due to intermittent renewable generation causes random voltage fluctuations. Therefore, probability distribution is needed to quantify voltage change under such stochastic scenarios. Here, we implement the first step of the proposed approach, i.e., derive the probability distributions of voltage change at any observation node due to random power change at a single actor node as well as due to the aggregate effect of all actor nodes. Let $\Delta S_{A}^{a}$ be change in complex power at phase $a$ of the actor node A. Then, using eqn. (\ref{eq:1}), the voltage change at phase-$a$ of an observation node $O$ can be expressed as:
\begin{equation}
   \Delta V_{OA}^{a}=\Delta V_{OA}^{a,r} + j\Delta V_{OA}^{a,i},
   \label{eq:4}
\end{equation}
\begin{equation}
\begin{split}
\Delta V_{OA}^{a,r} = \sum_{h,u}^{}\frac{-1}{|V_{A}^{h}|}[\Delta P_{A}^{h}(R_{OA}^{u}cos(\omega_{A}) - X_{OA}^{u}sin (\omega_{A})) \\ 
+ \Delta Q_{A}^{h}(R_{OA}^{u}sin(\omega_{A}) + X_{OA}^{u} cos(\omega_{A})) ] ,\\
\Delta V_{OA}^{a,i} = \sum_{h,u}^{}\frac{-1}{|V_{A}^{h}|}[
\Delta P_{A}^{h}(R_{OA}^{u}sin(\omega_{A}) + X_{OA}^{u} cos(\omega_{A})) + \\
\Delta Q_{A}^{h}(X_{OA}^{u}sin (\omega_{A})-R_{OA}^{u}cos(\omega_{A}))]
\end{split}
\label{eq:5}
\end{equation}
where $h$ $\epsilon$ $\tilde{H}$ and $u$ $\epsilon$ $\tilde{U}$. The sets $\tilde{H}$ and  $\tilde{U}$ denote different phases (i.e., $a,b,c$) and different phase sequences (i.e., $aa,ab,ac$), respectively. $\Delta P_{A}^{h}$ and $ \Delta Q_{A}^{h}$ are the active and reactive power changes, respectively. $R_{OA}^{h}, X_{OA}^{h}$ are the resistance and reactance of shared path between the observation node $O$ and actor node $A $ from the source node. $V_{A}^{h} $ denotes the base voltage of actor node $A$. For brevity, the derivation is shown for phase $a$. However, the same steps can be followed with the corresponding phase quantities to derive expressions for other phases. The real and imaginary parts of voltage change can further be simplified as,
\begin{equation}
\begin{split}
\Delta V_{OA}^{a,r} = \Bigg[ C_{OA}^{aa,r}\Delta P_{A}^{a} + D_{OA}^{aa,r}\Delta Q_{A}^{a} + C_{OA}^{ab,r}\Delta P_{b}^{b} + \\
 D_{OA}^{ab,r}\Delta Q_{b}^{b}+ C_{OA}^{ac,r}\Delta P_{c}^{c} + D_{OA}^{ac,r}\Delta Q_{c}^{c} \Bigg] ,\\
\Delta V_{OA}^{a,i} = \Bigg[ C_{OA}^{aa,i}\Delta P_{A}^{a} + D_{OA}^{aa,i}\Delta Q_{A}^{a} + C_{OA}^{ab,i}\Delta P_{b}^{b} + \\
 D_{OA}^{ab,i}\Delta Q_{b}^{b} + C_{OA}^{ac,i}\Delta P_{c}^{c} + D_{OA}^{ac,i}\Delta Q_{c}^{c} \Bigg] ,
\end{split}
\label{eq:6}
\end{equation}
Eqn. (\ref{eq:6}) is further written in compact form as,
\begin{equation}
\begin{split}
\Delta V_{OA}^{a,r} = \bm{ {C_{OA}^{a,r}}^{\tr} \Delta S_{A} }, \\
\Delta V_{OA}^{a,i} = \bm{ {C_{OA}^{a,i}}^{\tr} \Delta S_{A} },
\end{split}
\label{eq:7}
\end{equation}
where,
\begin{equation*}
\begin{split}
\bm{ {C_{OA}^{a,r}}^{\tr}} = [C_{OA}^{aa,r} \hspace{0.15cm} D_{OA}^{aa,r} \hspace{0.15cm} C_{OA}^{ab,r} \hspace{0.15cm} D_{OA}^{ab,r} \hspace{0.15cm} C_{OA}^{ac,r} \hspace{0.15cm} D_{OA}^{ac,r} ] \\
\bm{ {C_{OA}^{a,i}}^{\tr}} = [C_{OA}^{aa,i} \hspace{0.15cm} D_{OA}^{aa,i} \hspace{0.15cm} C_{OA}^{ab,i} \hspace{0.15cm} D_{OA}^{ab,i} \hspace{0.15cm} C_{OA}^{ac,i} \hspace{0.15cm} D_{OA}^{ac,i} ]
\end{split}
\end{equation*}	
The constants $C$ and $D$ are the functions of the line impedances and base voltages as explained in eqn. (\ref{eq:4}). For brevity, the exact expressions are omitted from here and are provided in the Appendix A. The power change vector is written as,
\begin{equation*}
\begin{split}
\bm{\Delta S_{A}} = [ \Delta P_{A}^{a}, \hspace{0.15cm}  \Delta Q_{A}^{a}  \hspace{0.15cm} , \Delta P_{A}^{b} \hspace{0.15cm}  , \Delta Q_{A}^{b} \hspace{0.15cm}, \Delta P_{A}^{c} \hspace{0.15cm}  ,\Delta Q_{A}^{c}]^\text{T} \\
\end{split}
\end{equation*}	
Thus, because of random power changes at each actor node, the power change vector $\bm{\Delta S_{A}}$ is a random vector with mean vector $\bm{\mu_{\Delta S_{A}}}$ and covariance matrix $\textstyle\sum_{\Delta S_{A}}$. 
The covariance matrix $\textstyle\sum_{\Delta S_{A}}$ quantifies the correlation of power changes among various phases of a particular actor node $A$. The diagonal elements denote variances of power change at each phase and off-diagonal elements contain the correlation between the power changes. Thus, it can be seen from (\ref{eq:7}) that the voltage change at an observation node $O$ due to actor node $A$ is the weighted combination of random vector $\bm{\Delta S_{A}}$, where weights (i.e., $\bm{ {C_{OA}^{a,r}}^{\tr}}$ and $\bm{ {C_{OA}^{a,i}}^{\tr}}$) are constant terms. Invoking the Lindeberg-Feller central limit theorem, it can be shown that the $\Delta V_{OA}^{a,r}$ and $\Delta V_{OA}^{a,i}$ converges in distribution to a Gaussian random variables, i.e.,  
\begin{equation}
\small
\begin{split}
\Delta V_{OA}^{a, r} \overset{D}{\rightarrow} \mathcal{N} (\mu_{OA}^{a,r}=\bm{ {C_{OA}^{a,r}}^{\tr}} \bm{\mu_{\Delta S_{A}}} ,  {\sigma_{OA}^{a,r}}^{2}=\bm{ {C_{OA}^{a,r}}^{\tr}} \textstyle\sum_{\Delta S_{A}} \bm{ {C_{OA}^{a,r}}} )\\
\Delta V_{OA}^{a, i} \overset{D}{\sim} \mathcal{N} (\mu_{OA}^{a,i}=\bm{ {C_{OA}^{a,i}}^{\tr}} \bm{\mu_{\Delta S_{A}}},  {\sigma_{OA}^{a,i}}^{2}=\bm{ {C_{OA}^{a,i}}^{\tr}} \textstyle\sum_{\Delta S_{A}} \bm{ {C_{OA}^{a,i}}} ),
\end{split}
\label{eq:8}
\end{equation}
where, $\pmb{\mu_{OA}^{a,r}}$ and $\pmb{\mu_{OA}^{a,i}}$ are the mean vectors of real and imaginary parts of voltage change, respectively.  ${\sigma_{OA}^{a,r}}^{2}$ and ${\sigma_{OA}^{a,i}}^{2}$ are the variances of real and imaginary parts of voltage change, respectively. For investigating the relationship between the real and imaginary parts of the voltage change, a new bivariate random vector is defined,
\begin{equation}
\begin{split}\begin{bmatrix}
\Delta V_{OA}^{a,r} \\ \Delta V_{OA}^{a,i}
\end{bmatrix} \sim \mathcal{N} \Bigg[\begin{bmatrix}
\mu_{OA}^{a,r} \\
\mu_{OA}^{a,i}
\end{bmatrix}, \begin{bmatrix}
{\sigma_{OA}^{a,r}}^{2} \hspace{0.5cm} k_{OA}^{a}   \\
k_{OA}^{a}  \hspace{0.5cm} {\sigma_{OA}^{a,i}}^{2}  
\end{bmatrix} \Bigg],
\end{split}
\label{eq:9}
\end{equation}
where, $k_{OA}^{a}= \bm{ {C_{OA}^{a,r}}^{\tr}} \textstyle\sum_{\Delta S_{A}} \bm{ {C_{OA}^{a,i}}} $ is the covariance between the real and imaginary parts of the voltage change due to single actor node $A$.
Eqn. (\ref{eq:9}) provides the probability distribution of voltage change at a particular observation node due to single actor node $A$. A similar approach can be used to compute individual voltage change distributions due to each actor node in the network. Now, we need to obtain the voltage change distribution due to the aggregate effect of all actor nodes. Using the superposition property in (\ref{eq:1}), the voltage change due to cumulative effect of power changes at multiple actor nodes can be expressed as \cite{munikoti2020probabilistic},

\begin{equation}
\begin{split}
\Delta V_{O}^{a, r} =  \sum_{A=1}^{L} \Delta V_{OA}^{a, r},
\end{split}
\label{eq:10}
\end{equation}
where $L$ is the number of actor nodes. By leveraging (\ref{eq:7}), the net voltage change can be written as,
\begin{equation}
\begin{split}
\Delta V_{OA}^{a,r} & = \sum_{A=1}^{N} \underset{1 \times 6}{ \bm{ {C_{OA}^{a,r}}^{\tr}} 
\underset{6 \times 1}{\bm{\Delta S_{A} }}} \\
& = [\bm{ {C_{O1}^{a,r}}^{\tr} } \bm{ {C_{O2}^{a,r}}^{\tr} } \hdots \bm{ {C_{ON}^{a,r}}^{\tr} } ][\bm{\Delta S_{1} } \bm{\Delta S_{2} } \hdots \bm{\Delta S_{N} }]^{\tr} \\
& = \underset{1 \times 6N}{\bm{ {C_{O}^{a,r}}^{\tr} }} \underset{6N \times 1}{\bm{\Delta S }}
\end{split}
\label{eq:11}
\end{equation}
where $\bm{ {C_{O}^{a,r}}^{\tr} } $ and $ \bm{\Delta S }$ are the long vectors, composed of a constant term and the power change vector corresponding to each actor node, respectively. Similarly, the imaginary part of voltage change can be written as,
\begin{equation}
\begin{split}
\Delta V_{OA}^{a,i} & = \sum_{A=1}^{N} \underset{1 \times 6}{ \bm{ {C_{OA}^{a,i}}^{\tr}} 
\underset{6 \times 1}{\bm{\Delta S_{A} }}}=\underset{1 \times 6N}{\bm{ {C_{O}^{a,i}}^{\tr} }} \underset{6N \times 1}{\bm{\Delta S }} \\
\end{split}
\label{eq:12}
\end{equation}
The equations (\ref{eq:11}) and (\ref{eq:12}) possess a similar form as that of (\ref{eq:7}), i.e., the net voltage change is the weighted combination of power change vector $ \bm{\Delta S }$. Here, weight $\bm{ {C_{O}^{a,r}} } $ is a constant vector comprising of line impedances and node base voltages, whereas, $ \bm{\Delta S }$ comprises of power change at all phases of all actor nodes. Now, invoking the same Lindeberg-Feller central limit theorem, the real $\Delta V_{O}^{a,r}$ and imaginary $\Delta V_{O}^{a,i}$ part of aggregate voltage change can be shown to converge in distribution to a Gaussian random variables with the following parameters:
\begin{equation}
\begin{split}
\Delta V_{O}^{a, r} \overset{D}{\rightarrow} \mathcal{N} (\mu_{O}^{a,r}=\bm{ {C_{O}^{a,r}}^{\tr}}\bm{\mu_{\Delta S}}, {\sigma_{O}^{a,r}}^{2}= \bm{ {C_{O}^{a,r}}^{\tr}} \textstyle\sum_{\Delta S } \bm{ {C_{O}^{a,r}}}) \\
\Delta V_{O}^{a, i} \overset{D}{\rightarrow} \mathcal{N} (\mu_{O}^{a,i}=\bm{ {C_{O}^{a,i}}^{\tr}} \bm{\mu_{\Delta S }}, {\sigma_{O}^{a,i}}^{2}=\bm{ {C_{O}^{a,i}}^{\tr}} \textstyle\sum_{\Delta S} \bm{ {C_{O}^{a,i}}} )
\end{split}
\label{eq:13}
\end{equation}
Similar to the single actor node case, the correlations between the real and imaginary parts of net voltage change is captured by defining a new random vector as:
\begin{equation}
\begin{split}\begin{bmatrix}
\Delta V_{O}^{a,r} \\ \Delta V_{O}^{a,i}
\end{bmatrix} \sim \mathcal{N} \Bigg[\begin{bmatrix}
\mu_{O}^{a,r} \\
\mu_{O}^{a,i}
\end{bmatrix}, \begin{bmatrix}
{\sigma_{O}^{a,r}}^{2} \hspace{0.5cm} k_{O}^{a}   \\
k_{O}^{a}  \hspace{0.5cm} {\sigma_{O}^{a,i}}^{2}   
\end{bmatrix} \Bigg]
\end{split}
\label{eq:14}
\end{equation}
where, $k_{O}^{a}= \bm{ {C_{O}^{a,r}}^{\tr}} \textstyle\sum_{\Delta S } \bm{ {C_{O}^{a,i}}} $ is the covariance between the real and imaginary parts of net voltage change. 
Equations (\ref{eq:9}) and (\ref{eq:14}) provide the probability distribution due to single actor node and the aggregation of multiple actor nodes, respectively. The next sub-section focuses on computing the statistical distances between these distributions and presents the procedure to rank the actor nodes.

\subsection{Information theoretic metrics as DVI indicators}
This sub-section implements the second step of our proposed approach, i.e., calculate statistical distances between the probability distributions (derived in earlier sub-section), and rank the actor nodes based on the computed distances. The information theoretic distance metrics which are potential indicators of DVI nodes are defined first.

\subsubsection{Kullback-Liebler distance}
Kullback-Liebler (KL) distance quantifies how much one probability distribution differs from another probability distribution.
KL divergence between two multivariate Gaussian distributions ($\mathcal{N}_{0}$ and $\mathcal{N}_{1}$) of dimension $k$ with means ($\mu_{0}$ and $\mu_{1}$) and covariance matrices ($\bm{\textstyle \sum_{0}}$ and $\bm{\textstyle \sum_{1}}$) 
can be written as:
\begin{equation}
\begin{split}
D_{KL}(\mathcal{N}_{0} || \mathcal{N}_{1}) = \frac{1}{2}\Bigg[ tr(\bm{\textstyle \sum_{1}^{-1}} \bm{\textstyle \sum_{0}} \\
+ (\mu_{1} - \mu_{0})^{\tr} \bm{\textstyle \sum_{1}^{-1}}(\mu_{1} - \mu_{0}) -k + \ln\frac{|\bm{\textstyle \sum_{1}}|}{|\bm{\textstyle \sum_{0}}|} ) \Bigg]
\end{split}
\label{eq:15}
\end{equation}
where $tr(.)$ indicates trace of the matrix. Here, the KL distance between the distributions of voltage change at the observation
node due to change in power at an actor node $A$ ($\Delta V_{OA}$)
and due to change in power at all actor nodes ($\Delta V_{O}$) is given by 
\begin{equation}
\begin{split}
D_{KL}(\Delta V_{OA} || \Delta V_{O} ) = \frac{1}{2}\Bigg[ tr(\bm{\textstyle \sum_{\Delta V_{O}^{a}}^{-1}} \bm{\textstyle \sum_{\Delta V_{OA}^{a}}} \\
+ (\mu_{O}^{a} - \mu_{OA}^{a})^{\tr} \bm{\textstyle \sum_{1}^{-1}}(\mu_{O}^{a} - \mu_{OA}^{a}) -2 + \ln\frac{|\bm{\textstyle \sum_{\Delta V_{O}^{a}}}|}{|\bm{\textstyle \sum_{\Delta V_{OA}^{a}}}|} ) \Bigg],
\end{split}
\label{eq:16}
\end{equation}
where $\bm{\textstyle \sum_{\Delta V_{OA}^{a}}}$ and $\bm{\textstyle \sum_{\Delta V_{O}^{a}}}$ are the covariances of $\Delta V_{OA}$ and $\Delta V_{O}$, respectively.

\subsubsection{Bhattacharyya distance}
Bhattacharyya (BC) distance measures the similarity of two probability distributions. It is related to the Bhattacharyya coefficient which is a measure of the amount of overlap between two statistical samples. BC distance between the distributions of voltage change at the observation node due to change in power at an actor node $A$ ($\Delta V_{OA}$) and due to change in power at all actor nodes ($\Delta V_{O}$) can be expressed as:
\begin{equation}
\begin{split}
D_{BC}(\Delta V_{OA} || \Delta V_{O} ) = \frac{1}{8}(\mu_{O}^{a} - \mu_{OA}^{a})^{\tr}\bm{\textstyle \sum} (\mu_{O}^{a} - \mu_{OA}^{a}) \\
+ \frac{1}{2} \ln(\frac{|\bm{\textstyle \sum}|} {\sqrt{|\bm{\textstyle \sum_{\Delta V_{O}^{a}}}||\bm{\textstyle \sum_{\Delta V_{OA}^{a}}}|} }) 
\end{split}
\label{eq:19}
\end{equation}
where $\bm{\textstyle \sum} = \frac{ \bm{\textstyle \sum_{\Delta V_{O}^{a}}}+ \bm{\textstyle \sum_{\Delta V_{OA}^{a}}}}{2}$ \\
\vspace{-0.5cm}
\subsection{Voltage Influencing score (VIS)}
The influence of an actor node on an observation node needs to be quantified for identifying its rank among all actor nodes. In this regard, we devise a novel index to quantify  
voltage influencing capacity, i.e., Voltage influencing score (VIS). For a given network scenario, i.e., the location of actor nodes with the variance of power change from historical data, the distances between voltage change distributions at an observation node due to each actor node and due to aggregate effect of all actor nodes are used to rank the actor nodes. Here, these distances are employed to compute the VIS between any pair of observation and actor node as:
\begin{equation}
    VIS(O,A) = \frac{\frac{1}{D(A,O)}-\frac{1}{D(1,O)}}{\frac{1}{D(A',O)}-\frac{1}{D(1,O)}},
    \label{eq:23}
\end{equation}
where $D(A,O)$ is the statistical distance between the voltage change distribution at an observation node $O$ due to aggregate effect of all actor nodes and when actor node $A$ is solely present in the system. The distance can be computed with any of the metrics described in the earlier subsection. For an observation node $O$, the lower the distance, the more the actor node $A$ contributes to aggregate voltage change and consequently the more influencing the actor node is and vice-versa. Therefore, the VIS is expressed in terms of inverse of distance. To provide an absolute sense to the score, VIS is normalized with minimum and maximum values. For a particular observation node, the ideal location of actor node $A'$ for minimum distance would be the same observation node location. 
On the other hand, the maximum distance location would always be the source node as it has minimum influence on voltage fluctuations of any observation node. These minimum and maximum distances are used to normalize VIS as shown in eqn. (\ref{eq:23}).

In the final step of the proposed approach, VIS is leveraged for ranking the actor nodes. The nodes are ranked in an ascending order based on VIS. The topmost (Rank 1) and bottom-most (Rank $L$ for the case of $L$ actor nodes) ranks will be assigned to the actor
nodes in the following way:
\begin{equation}
\begin{split}
\text{Rank} \hspace{0.15cm} 1 : \argmaxA_A VIS(O , A) \\
\text{Rank} \hspace{0.15cm} L : \argminA_A VIS(O, A) \\
\end{split}
\label{eq:20}
\end{equation}
It is worth to note that the proposed approach can work with any other information theoretic metric (Frechet distance, Jensen-Shannon distance, etc.), and for illustration purposes, the method is evaluated with two metrics (KL divergence \& Bhattacharyya distance) in the results section.
\vspace{-0.8cm}
\section{Results and Discussion}
\vspace{-0.4cm}
The efficacy of the proposed method in identifying dominant voltage influencer nodes is evaluated in this section. The baseline method is the conventional Monte-Carlo simulation-based approach, and both algorithms are implemented in the standard IEEE 37-node test system. This test network is chosen as it denotes a typical unbalanced distribution system and is used by many researchers for illustrating the efficiency of their methods \cite{munikoti2020analytical, munikoti2020probabilistic}. The nominal voltage of the test system is $4.8$ kV. A scenario is generated with $15$ actor nodes distributed randomly in the IEEE 37-node network. Change in real and
reactive power at $15$ actor nodes is modeled as zero-mean Gaussian random vector. It is worth noting that the proposed approach is generic for any choice of actor node and power change distributions, and the simulated case-study is merely a way to illustrate its performance. Three different PV sizes are considered in this case study. The mean and variance of real and reactive power change of all three kinds of PV capacities along with their location and phase information are as follows: 
\begin{equation}
\begin{split}
\bm{\Delta S_{A}} \sim \mathcal{N} \Bigg( \begin{bmatrix}
0 \\ 
0
\end{bmatrix} , \begin{bmatrix}
1.5 \hspace{0.4cm} -0.05 \\ 
-0.05 \hspace{0.4cm} 0.25
\end{bmatrix} \Bigg), \\ 
A \hspace{0.1cm} \epsilon \hspace{0.1cm}  \{7^{c},8^{a},12^{c},27^{b},28^{c} \} \\
\bm{\Delta S_{A}} \sim \mathcal{N} \Bigg( \begin{bmatrix}
0 \\ 
0
\end{bmatrix} , \begin{bmatrix}
3 \hspace{0.4cm} -0.1 \\ 
-0.1 \hspace{0.4cm} 0.5
\end{bmatrix} \Bigg), \\
A \hspace{0.1cm} \epsilon \hspace{0.1cm}  \{7^{b},18^{a},22^{c},34^{b},36^{b} \}   \\
\bm{\Delta S_{A}} \sim \mathcal{N} \Bigg( \begin{bmatrix}
0 \\ 
0
\end{bmatrix} , \begin{bmatrix}
4.5 \hspace{0.4cm} -0.2 \\ 
-0.2 \hspace{0.4cm} 0.75
\end{bmatrix} \Bigg), \\
A \hspace{0.1cm} \epsilon \hspace{0.1cm}  \{9^{c},14^{c},26^{c},30^{a},31^{a}\}
\end{split}
\label{eq:21}
\end{equation}
where superscript over actor nodes, i.e., $\{a,b,c\}$ represent respective phases of actor nodes at which power is varying. 
The change in power across different actor nodes can be correlated because of environmental factors. The DERs such as PVs and wind turbines are expected to exhibit a similar generation profile due to their geographical proximity. Further, the real and reactive power of inverter-based DERs is negatively
correlated. The underlying covariance structure $\textstyle\sum_{\Delta S}$ can be estimated based on historical or irradiance-related data. The base loads on the test network are the same as mentioned in the IEEE PES Distribution system analysis subcommittee report. For actor nodes, the variance of change in real power and reactive power are present in the diagonal elements of covariance matrices as shown in eqn. (\ref{eq:21}). The off-diagonal elements of the covariance matrices capture the covariance between real and reactive power. These covariance of each actor nodes are combined to form the covariance matrix of power change vector $\textstyle\sum_{\Delta S}$ which corresponds to all actor nodes. Here, the correlation coefficient between $\Delta P$'s and $\Delta Q$'s for different actor nodes within the same phase is kept same and covariance between cross-phase terms is assumed to be zero. However, the proposed approach is quite general to accommodate other covariance structures as well. The variance of nodes other than the actor nodes is set to zero. 
\begin{table}[h!]
    \begin{minipage}{.5\linewidth}
      \caption{Observation node 7}
      \label{tab:1}
      \centering
\begin{tabular}{|c|c|c|c|} 
\hline
\multicolumn{1}{|l|}{\textbf{Node}}  & \multicolumn{1}{l|}{\textbf{MC}} & \multicolumn{1}{l|}{\textbf{KL}} & \multicolumn{1}{l|}{\textbf{BC}}  \\ 
\hline
7                                   & 1                                & 1                       & 1                                 \\ 
\hline
9                                & 0.577                            & 0.163                            & 0.129                             \\ 
\hline
12                                & 0.543                            & 0.150                            & 0.117                             \\ 
\hline
22                                   & 0.530                            & 0.127                            & 0.096                             \\ 
\hline
14                                 & 0.516                            & 0.137                            & 0.106                             \\ 
\hline
26                                  & 0.441                            & 0.092                            & 0.067                             \\ 
\hline
28                                 & 0.350                            & 0.084                            & 0.060                             \\ 
\hline
8                                  & 0.042                            & 0.065                            & 0.045                             \\ 
\hline
17                                   & 0.026                            & 0.050                            & 0.033                             \\ 
\hline
27                                 & 0.001                            & 0.043                            & 0.029                             \\
\hline
\end{tabular}
    \end{minipage}%
    \begin{minipage}{.5\linewidth}
      \centering
        \caption{Observation node 16}
        \label{tab:1b}
\begin{tabular}{|c|c|c|c|c|} 
\hline
\multicolumn{1}{|l|}{\textbf{Node}} & \multicolumn{1}{l|}{\textbf{MC}} & \multicolumn{1}{l|}{\textbf{KL}} & \multicolumn{1}{l|}{\textbf{BC}}  \\ 
\hline
14                                  & 1                                & 1                                & 1                                 \\ 
\hline
22                                      & 0.919                            & 0.840                            & 0.808                             \\ 
\hline
18                                     & 0.782                            & 0.170                          & 0.133                             \\ 
\hline
12                                    & 0.613                            & 0.576                            & 0.533                            \\ 
\hline
9                                     & 0.426                            & 0.422                           & 0.373                            \\ 
\hline
7                                    & 0.209                            & 0.290                            & 0.242                             \\ 
\hline
17                                  & 0.146                            & 0.230                            & 0.186                             \\ 
\hline
8                                     & 0.020                            & 0.111                            & 0.083                             \\ 
\hline
26                                   & 0.001                            & 0.176                            & 0.138                             \\ 
\hline
28                                & 0.001                            & 0.164                            & 0.128  \\                       
\hline
\end{tabular}
    \end{minipage} 
\end{table}
\vspace{-0.5cm}
\subsection{VIS for ranking DVI nodes }
To assess the performance of the proposed approach in identifying DVI nodes, VIS for two arbitrary observation nodes (i.e., $7$, $16$) are computed using two different distance metrics namely KL divergence (eqn. (\ref{eq:15})) and Bhattacharyya distance (eqn. (\ref{eq:16})). 
The results obtained by using the proposed approach is validated against ground truth values. The baseline approach to compute true DVI nodes is based on Monte-Carlo simulations (MCS) of load flow as explained in Section \RomanNumeralCaps{2-}{A}. Initially, the variance of voltage change magnitude at a given observation node is computed due to the presence of all actor nodes. Multiple power change scenarios are simulated by running $100000$ MCS with load flow method. Then, the reduction in the variance of the magnitude of voltage change is determined by setting the variance to zero for each actor node sequentially. Again, for each actor node case, different MCS are executed. Finally, the actor nodes are ranked based on the reduction in variance they bought when the variance of power change at the corresponding actor node is set to zero. 
Tables \ref{tab:1} and \ref{tab:1b} tabulates the VIS for observation nodes $7$ and $16$, respectively. MC, KL and BC refers to baseline MCS approach, proposed KL divergence and Bhattacharyya distance based methods, respectively. Node in the tables refers to actor node. It can be inferred that the VIS for the most dominant actor node is $1$ and it decreases as we move to lower rank nodes. For observation node-7 with KL distance metric, there is a small difference in VIS between nodes $22$ and $14$ implying that they exhibit almost equal voltage influencing capacity. However, there is a considerable difference when it comes to nodes $14$ and $26$. Thus, VIS allows us to quantitatively differentiate between dominant influencer nodes.


\subsection{Ranking of DVI nodes with VIS}
DVI nodes are identified by ranking nodes based on VIS.
Table \ref{tab:2} tabulate the top $10$ dominant voltage influencer nodes for observation nodes $7$ and $16$, respectively.  
\begin{table}[h!]
	\centering
	\caption{DVI nodes for observation nodes 7 and 16}
\begin{tabular}{|l|l|l|l|l|l|l|l|l|l|l|}
\hline
\textbf{\begin{tabular}[c]{@{}l@{}} Rank $\rightarrow$ \\ Metric 	$\downarrow$
\end{tabular}} & \textbf{1} & \textbf{2} & \textbf{3} & \textbf{4} & \textbf{5} & \textbf{6} & \textbf{7} & \textbf{8} & \textbf{9} & \textbf{10} \\ \hline
\multicolumn{11}{|c|}{\textbf{Observation node -7}} \\ \hline
\textbf{MC}                                                             & 7          & 9         & 12         & 22         & 14         & 26         & 28          & 8          & 17          & 27          \\ \hline
\textbf{KL}                                                                  & 7          & 9         & 12         & 14         & 22         & 26         & 28          & 8          & 17          & 18         \\ \hline
\textbf{BC}                                                                  & 7          & 9         & 12         & 14         & 22         & 26         & 28          & 8          & 17          & 18         \\ \hline
\multicolumn{11}{|c|}{\textbf{Observation node -16}} \\ \hline
\textbf{MC}                                                                  & 14          & 22         & 18         & 12         & 9         & 7         & 17          & 8          & 26          & 28          \\ \hline
\textbf{KL}                                                               & 14         & 22         & 12             & 9         & 7         & 17     & 26          & 18          & 28     & 8                   \\ \hline
\textbf{BC}                                                                 & 14         & 22         & 12             & 9         & 7         & 17     & 26          & 18          & 28     & 8                   \\ \hline
\end{tabular}
	\label{tab:2}
\end{table}	


According to Table \ref{tab:2}, the MC method indicate that the actor nodes $7$ and $14$ are Rank-1 nodes for observation node $7$ and $16$, respectively. The assignment of Rank-1 nodes is carried out correctly by both the metrics of the proposed approach. The reasons for rank-1 allocation are (1) the high variance of power change and (2) proximity of actor nodes to the given observation node in the IEEE 37-node test system. 

Apart from rank-1, utilities might be interested in identifying other dominant nodes which are next to rank-1. In this regard, we use Top-N accuracy, which can be defined as,
\begin{equation}
\small
   \frac{|\{\text{Predicted Top-N nodes}\} \cap \{\text{True Top-N nodes\}}|}{ N},
    \label{eq:22}
\end{equation}
where $N$ is the desired number based on the applications. The Top-5 and Top-10 accuracy results are presented for observation nodes $7$ and $16$ in Table \ref{tab:4}. The last column of the table denotes the mean accuracy when all the nodes of the IEEE-37 node network act as observation nodes. It can be observed that distance metrics KL and BC have fairly good identification accuracy, and the mean accuracy is more than $90\%$. Apart from Rank-1 node, our approach also provides the correct sequence of actor nodes when compared to baseline MC approach. However, in some positions, the order of actor nodes is flipped or offset by one or two units. This is because, the corresponding actor nodes have an equal influencing capacity for that particular observation node. For instance, in the case of observation node $7$, the position of actor nodes $22$ and $14$ are flipped. However, this is not a major concern here, as we are more interested in correctly identifying nodes lying in a particular band (i.e., Top-5, Top-10) rather than their exact order. The power at the Top-5 or Top-10 actor nodes can be efficiently regulated to mitigate voltage violations.  
\begin{table}[h!]
	\centering
	\caption{Identification accuracy of DVI nodes}
\begin{tabular}{|c|c|c|c|}
\hline
\textbf{\begin{tabular}[c]{@{}c@{}}Node $\rightarrow$\\ Accuracy $\downarrow$ \end{tabular}} & \textbf{\begin{tabular}[c]{@{}c@{}}obs. node-7\\ (\%)\end{tabular}} & \textbf{\begin{tabular}[c]{@{}c@{}}obs. node-16 \\ (\%)\end{tabular}} & \textbf{\begin{tabular}[c]{@{}c@{}}Mean \\ all nodes (\%)\end{tabular}} \\ \hline
\textbf{Top-5}                                                   & 100                                                                 & 80                                                                   & 91                                                                      \\ \hline
\textbf{Top-10}                                                  & 90                                                                  & 100                                                                  & 93                                                                      \\ \hline
\end{tabular}
	\label{tab:4}
\end{table}	

Furthermore, to get the overall influencing capacity of actor nodes, one can compute the mean VIS. For each actor node, the mean of VIS is taken across all observation nodes of the network. It can be observed From Table \ref{tab:5} that actor nodes $17$ and $30$ have maximum and minimum influencing capacity among all the actor nodes present in the network. This assignment is because of their topological position in the distribution network and the magnitude of power variance due to associated DER units or load variation.

\begin{table}[h!]
	\centering
	\caption{Average VIS of actor nodes}
	\begin{tabular}{ |c|c|c|c|c|c|} 
		\hline
		\textbf{Nodes} & \textbf{VIS} & \textbf{Nodes} & \textbf{VIS} & \textbf{Nodes} & \textbf{VIS} \\ 
		\hline
		7 & 0.220 & 17 & 0.444 & 28 & 0.156 \\ 
		\hline
		8 & 0.220 & 18 & 0.304 & 30 & 0.154 \\ 
		\hline 
		9 & 0.212 & 22 & 0.349 & 31 & 0.189 \\ 
		\hline 
		12 & 0.258 & 26 & 0.163 & 34 & 0.378 \\ 
		\hline 
		14 & 0.300 & 27 & 0.166 & 36 & 0.427 \\ 
		\hline 
	\end{tabular}
	\label{tab:5}
\end{table}

\begin{table}[h!]
	\centering
	\caption{Running time of various approaches}
\begin{tabular}{|c|c|}
\hline
\textbf{\begin{tabular}[c]{@{}c@{}}Time\\ Metric\end{tabular}} & \textbf{\begin{tabular}[c]{@{}c@{}}Running time\\ Top-5   (s)\end{tabular}} \\ \hline
\textbf{MC}                                                    & 12039                                                                       \\ \hline
\textbf{KL}                                                    & 0.92                                                                        \\ \hline
\textbf{BC}                                                    & 0.96                                                                        \\ \hline
\end{tabular}
	\label{tab:6}
\end{table}	
\vspace{-0.5cm}
\subsection{Computational complexity of proposed method}
Apart from correctly identifying the dominant voltage influencing nodes, the proposed approach offers a considerable computational advantage over the conventional approaches.
Table \ref{tab:6} reports the execution time of various approaches to identify Top-5 actor nodes for the observation node $7$. It can be seen that the proposed approach involving any of the two distance metrics is multiple order faster than that of the conventional approach, which takes around $3.2$ hrs. All the experiments are conducted on a machine with Intel i7 processor running at 2.2 GHz. This results demonstrates the computational advantage of the proposed approach over conventional methods.

The proposed approach of identifying DVI nodes can enhance voltage control strategies in multiple ways. One of the approaches is based on system partitioning, where a set of effective actor nodes is used for control actions. Our approach helps to identify such set of actor nodes in a computationally efficient manner. Given a particular scenario of power change in the system, one can determine the set of common observation nodes for which a particular actor node induces the maximum impact on voltage profiles. Therefore, the set of common observation nodes can form a cluster and it is cost effective to control power variations on the most dominant actor node to quickly restore voltages to their safe operational limits within that cluster. Along with clustering, our method of ranking nodes can also be used for optimal allocations of DERs or fault current limiters which involves sensitivity analysis. These control and asset allocation strategies will be investigated as part of our future work. 
\vspace{-0.2cm}
\section{Conclusion and Future work}
The conventional methods to identify dominant voltage influencer nodes (DVI) are Monte-Carlo simulation-based that are computationally expensive. Therefore, this work focuses on an analytical approach and proposes statistical distance metrics to devise VIS that quantifies voltage influencing capacity of actor nodes. The VIS is then leveraged to identify the DVI nodes. The proposed framework computes VIS not solely on the basis of correlation between change in power at the actor node and change in voltage at the observation node, but it also relies on the magnitude of voltage variation that is being caused due to power change at actor nodes. Effectiveness and computational efficiency
of the proposed method are illustrated by comparing
the results with conventional method of identifying DVI nodes using
Monte-Carlo simulations in IEEE 37-node test system. From the results, it can be inferred that the proposed statistical distance metrics effectively predict the DVI nodes while substantially reducing the execution time. This scalable approach of identifying DVI nodes can find applications in a variety of areas such as in utilizing DVI nodes for voltage control at strategic locations, identifying most influencer DERs that are responsible for voltage fluctuations, among others. Employing DVI nodes for network partitioning and efficient control applications will be pursued as part of our future work.

\section*{Acknowledgment}
This material is based upon work partly supported by the Department of Energy, Office of Energy Efficiency and Renewable Energy (EERE), Solar Energy Technologies Office, under Award \# DE-EE0008767 and National science foundation under award \# 1855216.

\bibliographystyle{IEEEtran}
\bibliography{IEEEabrv, DIVF_3ph}

\end{document}